\documentclass[12pt,preprint]{aastex}
\usepackage{natbib}
\usepackage{amssymb, amsmath, amsbsy, epsfig, epsf, slashbox, rotating}
\usepackage{color}
\slugcomment{Accepted by {\it The Astrophysical Journal Letters}}
\begin{document}

\title{Probing Population III Stars in Galaxy IOK-1 at $z=6.96$ through He II Emission}
\author{Zheng Cai \altaffilmark{1}, Xiaohui Fan, Linhua Jiang, Fuyan Bian, Ian McGreer, Romeel Dav\'e, Eiichi Egami, Ann Zabludoff}
\affil{Steward Observatory, University of Arizona, Tucson, AZ 85721}
\author{Yujin Yang}
\affil{Max-Planck-Institut f${\ddot{\rm{u}}}$r Astronomie, K$\ddot{\rm{o}}$nigstuhl 17, D- 69117, Heidelberg, Germany}
\author{S. Peng Oh}
\affil{Department of Physics, University of California, Santa Barbara, CA 93106}

\altaffiltext{1} {Email: caiz at email.arizona.edu}
%
%

\begin{abstract}
The He II $\lambda 1640$ emission line has been suggested as a direct
probe of { Population III (Pop III)} stars at high-redshift,
since it can arise from highly energetic ionizing photons associated
with hot, metal free stars.  We use the HST WFC3/F130N IR narrowband filter
to { probe} He II $\lambda 1640$ emission in galaxy IOK-1 at $z=6.96$. The sensitivity of this measurement is $\gtrsim$5$\times$ deeper than for previous
measurements.  From this deep narrowband imaging, combined with
broadband observations in the F125W and F160W filters,
we find the He II flux to be $1.2\pm
1.0\times 10^{-18}\ \rm{ergs}\ \rm{s}^{-1}\ \rm{cm}^{-2}$,
corresponding to a 1$\sigma$ upper limit on the Pop III star formation rate (SFR) of
$\sim 0.5\ \rm{M}_{\odot}\ \rm{yr}^{-1}$  for the case of a Salpeter IMF
with $50\lesssim M/M_{\odot}\lesssim 1000$ and mass loss.  Given that the
broadband measurements can
be fit with a UV continuum spectral flux density of $\sim 4.85
\times10^{-10} \times \lambda^{-2.46}\ \rm{ergs}\ \rm{s}^{-1}\
\rm{cm}^{-2}$\ \AA$^{-1}$, which corresponds to an overall SFR of $\sim 16^{+2.6}_{-2.6}$  M$_{\odot}\ \rm{yr}^{-1}$, massive Pop III stars represent
$\lesssim 6$\% of the total star formation.  This measurement
places the strongest limit yet on metal-free
star formation at high redshift, 
although the exact conversion from He II luminosity to Pop III SFR
is highly uncertain due to the unknown IMF, stellar evolution, and
photoionization effects.
Although we have not detected He II $\lambda1640$ at more than the 1.2$\sigma$ level, our work
suggests that a $\gtrsim 3\sigma$ level detection is possible with JWST.

\end{abstract}

\section{Introduction}

The first stars, i.e., Population III stars with zero metallicity, are likely to be very massive, due to inefficient cooling and the large Jeans mass in the early Universe \citep{abel2002, yoshida2006}. 
Massive Pop III stars are probably key contributors both of the ionization photons that caused cosmic reionization and to the early phases of cosmic chemical evolution \citep{whalen08, mackey03, bromm03}. Although recent observations have provided major breakthroughs in understanding the universe at $z\gtrsim 7$ \cite[e.g.,][]{bouwens10_z8, yan11}, detection of 
high redshift galaxies that contain Pop III stars remains a major challenge observationally. 
Indirect constraints on Pop III stars come from mapping the reionization history \citep*{fan06} or from studying the chemical evolution and the enrichment history through the observation of Extremely Metal-Poor stars and Hyper Metal-Poor stars \citep{beers2005, bessel04}. 
{ There are still no direct dectections of high redshift galaxies hosting { star formation in zero metallicity gas}.} 

Models show that Pop III stars are considerably hotter than present-day stars, for a given stellar mass \cite[e.g.,][]{johnson10, bromm03}.  
Emission of ionizing radiation, specifically photons with energies above 54.4eV that ionize He II, is enhanced by the high surface temperature photospheres of Pop III stars.  He II $\lambda$1640 has been long suggested as a direct probe of Pop III stars \citep{tumlinson00, oh01, schaerer02}. HeII $\lambda$ 1640 emission is also relatively easy to model compared to resonance lines such as Ly$\alpha$ or He II $\lambda304$. The optical depth to He II $\lambda1640$ is small in almost all circumstances, so it does not suffer the radiative transfer effects that complicate the interpretation of Ly$\alpha$ flux in high-redshift galaxies \citep{schaerer03}. In addition, it does not suffer from IGM Gunn-Peterson absorption as  Ly$\alpha$ does \citep{tumlinson00}. 
 
{How can He II $\lambda1640$ emission from star formation in zero metallicity gas be distinguished from other sources ? Calculations show that the He II $\lambda$1640 emission in a metal-poor stellar population with metallicity as low as Z$\sim$10$^{-3}$ is still $<$10$^{-3}$ times lower than that in a population formed in zero metallicity gas \citep{tumlinson03, schaerer03}.}
 It is also known that nebular He II $\lambda4686$ was detected in{ a fairly large fraction of metal-poor HII regions} \cite[e.g.,][]{skillman93}. { The follow-up observations and theoretical investigations suggest that Wolf-Rayet (W-R) stars are responsible for this nebular He II emission  \citep{demello98}, but the hardness of this emission more than one order of magnitude weaker than that originated by massive Pop III stars \citep{schaerer03, jimenez06}. } Also, the He II $\lambda1640$ equivalent width of massive Pop III stars could be a factor of 2 larger than that of the typical AGN  \citep{prescott09, elvis94}. 

A number of studies have been carried out to search for He II emission in high-redshift galaxies at $4<z<6.6$.
\citet{dawson04}, \citet{ouchi08}, and \citet{nagao05} searched for He II $\lambda1640$ in either stacked or individual spectra of Ly$\alpha$ emitting galaxies.
\cite{nagao08} carried out a survey for Ly$\alpha$-He II emitters using a combination of intermediate and narrow-band filters in the optical window.
But these observations so far have yielded only nondetections, constraining the massive Pop III star formation rate (hereafter SFR$_{\rm{PopIII}}$) to a few M$_{\odot}\ \rm{yr}^{-1}$, usually a few tenths of the overall SFR in these galaxies. Yet \citet{jimenez06} predict SFR$_{\rm{PopIII}}$ still to be signficant at z $\sim$ 3-5. Even stronger metal-free star formation is expected at higher redshifts, due to less chemical feedback from Pop III stars after their initial burst. 
However, at $z>7$, \ion{He}{2} is in the J-band, making ground-based observations extremely challenging.
 { Due to its large field of view, high throughput, and low background, the new IR channel of the Wide Field Camera 3 (WFC 3) on HST is the most powerful tool for detecting galaxies at z$\gtrsim$ 7 and for studying galaxy evolution at the end of cosmic reionization. }
 

In this paper, we report the strongest upper limit yet on the contribution of Pop III stars in high-redshift galaxies, using deep narrow band imaging of galaxy
IOK-1 \citep{iye06} at $z=6.96$ in the HST/WFC3 F130N band, which includes He II $\lambda 1640$ at this redshift. 
We have also carried out deep observations in broad bands to measure the continuum level and  UV-based total SFR. 
In \S 2, we discuss the observations and data reduction. 
Photometric results, including the limit on He II flux, are presented in \S 3. 
We discuss the implication of our results on Pop III star formation in IOK-1 in \S 4. 
Throughout this paper, we adopt a cosmology based on the fifth year Wilkinson Microwave Anisotropy Probe (WMAP) data \citep{komatsu09}: $\Omega_{\Lambda}=0.72$, $\Omega_m= 0.28$, $\Omega_b=0.046$, and $\rm{H}_0=70\ \mbox{km}\ \rm{s}^{-1}\ \mbox{Mpc}^{-1}$. 

\section{Observation and the data reduction}
Our target, IOK-1, was discovered in a narrow band Ly$\alpha$ emitter survey using the NB973  filter on the Subaru Suprime-Cam (Iye et al. 2006), which is centered on Ly$\alpha$ at $z\simeq 7.02$. This galaxy is spectroscopically comfirmed at $z=6.96$.
IOK-1 has a total  Ly$\alpha$ luminosity of $1.1\pm0.2\times 10^{43}\ \rm{erg}\ \rm{s}^{-1}$, corresponding to a Ly$\alpha$-based SFR $\rm{SFR}_{Ly\alpha}$ $\sim10\pm 2\ \rm{M}_{\odot}\ \rm{yr}^{-1}$. (Iye et al. 2006). \footnotemark { 
\footnotetext[\value{footnote}]{ These error bars only reflect the photometric uncertainties. There are big systematic uncertainties due to Ly$\alpha$ radiative 
transfer effect, galaxy IMF and metallicity assumptions. The SFR$_{\rm{Ly}\alpha}$ could change 
by a factor of 2 given these systematic uncertainties (also see \S 4). } 


In the observed frame, the He II $\lambda 1640$ emitted by IOK-1 is located at $13054$ \AA, which is in the passband of the WFC3 F130N narrow band filter
($\lambda_0 = 13006\ \rm{\AA}$ and FWHM = 150\ \rm{\AA}).  Resonant scattering of Ly$\alpha$ photons and IGM absorption of the blue wing usually systematically redshift the Ly$\alpha$ line center up to  $\sim$ 700 km s$^{-1}$ against the systemic velocity \citep[e.g.,][]{steidel10}. Given that \ion{He}{2} will be located slightly blueward of the peak inferred from Ly$\alpha$ peak and have a narrower line width, we expect that the entire \ion{He}{2} line  falls securely into the most sensitive part of the F130N filter. 
IOK-1 was observed by HST WFC3 in March 2010. Eight orbits ($\sim 20,000\ \rm{sec}$ integration) were devoted to the F130N filter to measure the \ion{He}{2} flux;  we also
observed this galaxy using the F125W filter for two orbits to determine the underlying continuum level and UV continuum-based SFR. A F160W band observation was also carried out (Egami et al. 2011, in preparation).

The F130N observations are divided into two visits, each consisting of a 4-point dither sequence with one-orbit per dither. 
During the the first visit, the first two dither positions were affected by the presence of a bright ghost image very close to IOK-1.
We exclude these two images from further analysis. 
The F125W continuum observation was a two-orbit single visit, which is divided into two identical 4-point dither sequences, one per orbit. No ghost image affected
the F125W observations. 
The individual images were reduced by WFC3-IR standard pipelines. 
Both the F130N and F125W images were combined using Multidrizzle \citep{koekemoer02}
with $\rm{final}\_\rm{scale}=0\farcs06$, which is $0.48$ of the original pixel size, and  $\rm{final}\_\rm{pixfrac}=0.7$. 
High resolution images from the F130N and F125W bands are shown in Figure 1.

\section{Results}
\subsection{Photometry}
Photometry is performed with SExtractor \citep{bertin96} using the rms map converted from the inverse variance image generated by Multidrizzle \citep{casertano00}. We measure the fluxes in the broadband (F125W and F160W) and the narrowband (F130N) images using the same Kron-like elliptical aperture determined from the F130N image (black elliptical aperture in Figure 1). The results are listed in Table \ref{table:F130N_S}. We also measure the flux in the F160W image of IOK-1 (Egami et al. 2011, in preparation), using the same aperture as used in the F125W and F130N images, finding a flux density of $f^{\lambda}_{\rm{F160W}}$= $2.44 \pm0.14 \times 10^{-20}\ \rm{erg}\ \rm{s}^{-1}\ \rm{cm}^{-2}\ $\AA$^{-1}$, corresponding to a magnitude of  $m_{\rm{F160W}}$ = $25.98\pm0.06$. 
We fit the photometry by  assuming a model spectrum with a power law continuum with a dimensionless flux density $f_{con}= A (\lambda/1\rm{\AA})^{\beta}$, where A is a constant, and a narrow Gaussian for the \ion{He}{2} emission line of flux $\rm{F_{HeII}}$.
The best-fit SED and photometry are shown in Figure 2. 
We find:
\begin{equation}
\label{eq:spectra}
f_{con}(\lambda) = ( 4.85\pm 0.23 ) \times 10^{-10} (\lambda/1\ \rm{\AA})^{-2.46\pm0.36}, 
\end{equation}
\begin{equation}
\rm{F}_{\rm{HeII}}= (1.2 \pm 1.0) \times 10^{-18} \rm{ergs}\ \rm{s}^{-1}\ \rm{cm}^{-2}.
\end{equation}
At $z=6.96$, this corresponds to a total HeII luminosity of 
${L}_{\rm{HeII}}= \rm{F}_{\rm{HeII}}\times 4\pi D_L^2= (6.6$ $\pm 5.5)\times 10^{41} \rm{ergs}\ \rm{s}^{-1}$, 
or a 2-$\sigma$ upper limit of $1.11\times 10^{42}\ \rm{erg}\ \rm{s}^{-1}$.
This 2-$\sigma$ upper limit is a factor of $\gtrsim$4-5$\times$ deeper than previous measurements. 
The rest-frame equivalent width of \ion{He}{2} is $4.2\pm 3.5$\AA.  \footnotemark { 
\footnotetext[\value{footnote}]{ Note that our observations could only detect He II flux stronger than 3$\times$10$^{-18}\ \rm{erg}\ \rm{s}^{-1}\ \rm{cm}^{-2}$  at 3$\sigma$ level, corresponding to a He II $\lambda1640$ equivalent width of $\sim$11\AA, which is a factor of $\gtrsim$ 6 larger than that of normal stellar populations in the Lyman break spectrum of stacked z$\sim$3 galaxies \citep{shapley03}, and a factor of $\gtrsim$ 3 greater than the simulated maximum He II equivalent width generated by W-R stars \citep{brinchmann08}.}  } 
Our results also show that IOK-1 has a blue continuum. \citet{bouwens10} studied the value of the UV-continuum slope $\beta$ in the Hubble Ultra Deep Field (HUDF). For luminous $L^*_{z=3}$ galaxies \citep{steidel99}, $\beta\sim -2.0\pm 0.2$. For lower luminosity 0.1$L^*_{z=3}$ galaxies, $\beta \sim -3.0\pm 0.2$. 
IOK-1 has an absolute magnitude M$_{\rm{AB}} (1500\rm{\AA}) = - 21.3$, comparable to $L^*_{z=3}$. 

\begin{table}  
\caption{SExtractor photometry in different bands} 
\label{table:F130N_S}	
\centering 
\begin{tabular}{c c c} 
\hline\hline 
Filter &   Flux Density f$^\lambda$ & $\rm{m}_{\rm{AB}}$ \\
          &   ($10^{-20}\ 	\rm{erg}\ \rm{s}^{-1}\ \rm{cm}^{-2}\ \AA^{-1})$  & \\
\hline 
F130N &  $4.39 \pm 0.68$ & $25.42\pm0.17$ \\
\hline
F125W &  $4.07  \pm 0.19$ & $25.59\pm 0.05$\\
\hline
\end{tabular}
\end{table}

\subsection{Morphology}

The half-light radius $r_{1/2}$ of IOK-1, based on our SExtractor measurements \citep{bertin96} and corrected for WFC3 PSF broadening, is 0$\farcs$12 in F125W. 
 This corresponds to  $0.62\pm 0.04$ kpc.   IOK-1 is an extremly compact galaxy, consistent with the size of $L^*_{z=3}$ galaxies at z $\sim 7$ detected in the HUDF09 \citep{oesch10}. The observed surface brightness of IOK-1 is $\mu_{J} \sim 24.5\ \rm{mag}\ \rm{arcsec}^{-2}$, corresponding to a 
{surface brightness of $\mu_{rest}\sim 15.5\ \rm{mag}\ \rm{arcsec}^{-2} $, after (1+z)$^4$ correction for cosmological dimming. } 
From Figure 1, it is obvious that  IOK-1 consists of two components.
We use GALFIT \citep{peng03} to deblend the F125W continuum image, assuming an exponential profile for both components. The F125W image shows roughly equal brightness for the two components. The north west component has an effective radius of $0.49\pm 0.04$ kpc; and the southeast component has an effective radius of $0.57\pm0.03$ kpc. 
 The two components are projected about 0$\farcs$2 ($\sim$1 kpc) away from each other.
More detailed discussions  on the morphology are presented in Egami et al. (2011, in preparation). 
We also deblend the F130N narrow band image, which contains \ion{He}{2} $\lambda1640$ emission. The north-west component is somewhat brighter. However, given the low signal-to-noise ratio of the data, this difference, while intriguing, is not statistically significant. 

\section{Discussion}

What limit can our observations place on the SFR$_{\rm{PopIII}}$ and its contribution to the overall star formation in IOK-1 at $z \sim 7$ ? Photoionization by AGN or hot dense stellar winds from Wolf-Rayet stars \cite[see e.g.,][]{shapley03, brinchmann08} may contribute to \ion{He}{2} $\lambda1640$ emission although {at a lower level than metal-free stars.} More importantly, massive stars with the lowest, but non-zero metallicity ($<10^{-8}$) could generate a comparable amount of He II $\lambda 1640$ emission to Pop III stars \citep{schaerer03}.
However, here we assume that the observed \ion{He}{2} emission originates
entirely from metal-free stars. Therefore, our derived SFR$_{\rm{PopIII}}$ should be regarded as an upper limit. 
 In the case of constant star formation, at equilibrium, recombination line luminosities $L_l$ are proportional to the SFR$_{\rm{PopIII}}$ \citep{schaerer02}, so
\begin{eqnarray}
   L_{\rm{HeII}} &=& c_{1640} (1-f_{\rm{esc}}) Q(\rm{He}^+) \left( \frac{\rm{SFR}_{\rm{Pop III}}}{M_{\odot} \rm{yr}^{-1}}\right)\nonumber \\
   & = &L_{1640, \rm{norm}}\left( \frac{\rm{SFR}_{\rm{Pop III}}}{\rm{M}_{\odot} \rm{yr}^{-1}}\right) 
\end{eqnarray}
where $L_{1640, \rm{norm}}\equiv c_{1640}(1-f_{esc})Q(\rm{He}^+)$ { is the theoretical He II $\lambda 1640$ line luminosity normalised to SFR=1 M$_\odot$ yr$^{-1}$. $c_{1640}$ is the He II $\lambda1640$ emission coefficient given in Table 1 of \citet{schaerer03}: $c_{1640}=5.67\times 10^{-12}\  \rm{ergs}$ for $T_e=30k$K, where nebular emission is calculated assuming case B recombination.
 $Q(\rm{He}^+)$ is the number of $\rm{He}^+$ ionizing photons per second, $f_{esc}$ represents the fraction of total ionizing radiation released into the IGM without being coupled to the ISM in the galaxy. 
 We assume $f_{esc} = 0$, as it is expected to be small among high-redshift galaxies \cite[e.g.,][]{gnedin08}. 
 \citet{schaerer02} calculates the total $\rm{He}^+$ ionizing photon flux for a Salpeter IMF with a range of lower and upper mass cut-offs and different assumptions of mass loss.
\begin{table} [!h]
\centering 
\caption{Pop III star formation rate} 
\label{table:PopIII_SFR}	
\begin{tabular}{|c | c | c| c | } 
\hline\hline 
IMF(Salpeter) & Mass Loss & $L_{1640, \rm{norm}} $ & $\rm{SFR}_{\rm{Pop} III} $ \\  
                    &                & ($\mbox{ergs}\ \rm{s}^{-1}$) & $ (\rm{M}_{\odot}\ \rm{yr}^{-1})$ \\ 
\hline
$1\lesssim M/M_{\odot}\lesssim500$ & No &$ 9.66\times10^{40}$ & $ 6.8\pm4.7$\\ 
\hline 
$50\lesssim M/M_{\odot}\lesssim500$ & No & $6.01\times10^{41}$ & $ 1.1\pm0.9$\\ 
\hline
$1\lesssim M/M_{\odot}\lesssim500$ & Yes & $3.12\times10^{41} $ & $ 2.1\pm1.5$\\ 
\hline
$50\lesssim M/M_{\odot}\lesssim1000$ & Yes & $2.33\times10^{42}$ & $0.3\pm0.2$\\
\hline
\end{tabular}
\end{table}
 
 The conversion from He II $\lambda1640$ { luminosity} (L$_{\rm{HeII}}$) to SFR$_{\rm{PopIII}}$ is affected by a number of uncertainties, including:
 (1) IMF, (2) mass loss, and (3) photo-ionization model.   
 For the IMF, theoretical calculations show that 
the lack of efficient cooling mechanisms generally result in an extremely top-heavy IMF compared to a galactic IMF \citep{abel2002,  bromm04, yoshida2006, oshea06}. 
However, both the shape and mass range of such an IMF are poorly constrained.
\citet{scannapieco06} suggests a lower mass cutoff of 0.8 M$_{\odot}$ based on metal abundances in Galactic halo stars. Numerical simulations suggest that
stars as massive as  500 M$_{\odot}$  can be formed by accretion into a primodial protostar  \citep{bromm04}, which is generally taken as an upper limit  in different simulations \citep{schaerer02, scannapieco03, tumlinson06, raiter10}. 
Both a Salpeter IMF \citep{schaerer03, scannapieco03} and a Log-normal IMF \citep{tumlinson06} have been used in model calculations. 
Table 2 shows a factor of six difference in inferred SFR arising from different lower mass cutoffs in the Salpeter IMF.

{ As shown in Table 2, models with strong mass loss have a larger conversion factor $L_{1640, \rm{norm}}$; therefore, for the same HeII luminosity, they yield a lower SFR$_{\rm{PopIII}}$  \citep{schaerer03}. } \citet{kudritzki02} shows that very low-metallicity stars close to the Eddington limit are subject to non-zero mass loss. \citet{smith06} argue that massive shells around luminous blue variables and so-called supernova impostors indicate that continuum-driven winds or hydrodynamic explosions dominate the mass loss of very massive stars, a claim which is insensitive to metallicity and so could apply to Pop III stars.  
 Ekstrom (2007) explores the effects of rotation, anisotropic mass loss, and magnetic fields on the core size of a Pop III star, pointing out that, under certain conditions, very massive stars $(140M_{\odot}< M< 260M_{\odot})$ losing mass through rotation could avoid ending their lives as pair-instability supernovae (PISN). Whether Pop III stars end their lives as PISN or not will affect the metal ejection efficiency and hence affect the SFR$_{\rm{PopIII}}$. From Table \ref{table:PopIII_SFR}, stronger mass loss will lower $\rm{SFR}_{\rm{Pop III}}$ by a factor of more than three.

The derived SFR also depends on the photoionization model.  { \citet{schaerer02} calculates nebular emission lines using standard case B recombination, in which the optical depth of the HII region is large. Eq. (3) is derived under this assumption. However, for low metallicity nebulae ionized by very hot Pop III stars, the case B predictions for line and continuum emission may have non-negligible deviations from real nebular astrophysics \citep{raiter10}. Therefore, the SFR$_{\rm{PopIII}}$ upper limits derived from Eq. (3) then need to be qualitatively revised here. According to \citet{raiter10}, for a given He II luminosity, SFR$_{\rm{PopIII}}$ could be a few times higher, depending on the ionizing parameters, e.g., the hydrogen number density $n(H)$, inner radius of the nebula $r_{in}$, and hydrogen ionizing photon flux Q(H). }

{{ Our upper limit on the He II emission coombined with our detection of IOK-1 in the Spitzer IRAC 1 and 2 bands (Egami et al., in prep.) suggest that IOK-1 is dominated by stars of metallicity above the critical value (Z$_{\rm{crit}}\sim 5\times 10^{-4} \rm{Z}_{\odot}$), leading to a normal IMF (e.g., Bromm 2004).} 
Using the relation given in \citet{madau98}, the rest frame UV luminosity of the galaxy IOK-1 corresponds to an overall
$ \rm{SFR}\sim\ 20^{+0.9}_{-0.9}\ \rm{M}_{\odot}\ \rm{yr}^{-1}$}. { These error bars only reflect the uncertainties in the J-band magnitude. The true errors are larger as the conversion from UV luminosity to overall SFR is highly and systematically uncertain. In this conversion, Madau (1998) assumes solar metallicity; lower metallicities down to 0.0004 could lower the SFR by a factor of 2 \citep{schaerer03}. Furthermore, the conversion assumes a Salpeter IMF with mass ranging from 0.1 to 125M$_\odot$, whereas assuming a Scalo IMF would double the derived overall SFR. }
The UV-based overall SFR which should be less obscured is somewhat higher than that based on the Ly$\alpha$ emission line ($\sim 10\pm 2 \ \rm{M}_{\odot}\ \rm{yr}^{-1}$; \citealt{iye06}), but consistent given the
uncertainties in both the stellar population models and the Ly$\alpha$ radiative transfer effects. 

{ The conversion of UV luminosity to an overall SFR above assumes a conversion factor between UV luminosity and SFR for {Pop I and Pop II} stellar populations, without correcting for Pop III stars with a top-heavy IMF. Using the fiducial Pop III Salpeter IMF of 50 - 500 M$_{\odot}$ and no mass loss, we find that our derived upper limit of 
2.0 $\rm{M}_\odot\ \rm{yr}^{-1}$ for SFR$_{\rm{PopIII}}$ corresponds to an upper limit of $1.2\times 10^{-20}\ \rm{ergs\ s}^{-1} \rm{cm}^{-2}$ \AA$^{-1}$ in F125W flux density. 
Note that the observed flux density is $4.07\pm0.19 \times 10^{-20}\ \rm{ergs\ s}^{-1}  \rm{cm}^{-2}$ \AA$^{-1}$. After correcting for the UV-luminosity contributed by Pop III stars, the overall SFR is $\sim$ $16^{+2.6}_{-2.6}$ M$_\odot\ \rm{yr}^{-1}$. ({ As stated previously, this result could change a factor of 2 given the systematic uncertainties associated with the IMF and metallicity assumptions}). Thus, 
{IOK-1 is not dominated by very massive Pop III stars.} It should be noted that low mass metal-free stars do not produce an appreciable amount of He II $\lambda1640$ emission. Therefore, our observations only constrain the amount of high-mass Pop III star formation in the galaxy and cannot be used to rule out the existence of low mass, low surface temperature Pop III stars (although, as discussed in \S4, most theoretical calculations favor a top heavy IMF for Pop III). }  
  \citet{trenti09fIII} show that the fraction of Pop III stars $f_{\rm{III}}(z)$ per dark matter halo at $z\sim 10$ is only a few thousandths.  \citet{dave06} and \citet{tumlinson06} also suggest that $f_{\rm{III}}(z)$ in halos at $z\sim 7$ is small: $\lesssim 1\%$. {Taking into account the systemic uncertainties in the UV-based overall SFR, our observations indicate that  
 the galaxy IOK-1 cannot be dominated by Pop III stars with very top heavy IMFs. For example, for a Salpeter IMF with $50\lesssim M/M_{\odot}\lesssim 500$ and no mass loss, the ratio of SFR$_{\rm{PopIII}}$ to the overall SFR is $\lesssim$ 25\%. For a Salpeter IMF with $50\lesssim M/M_{\odot}\lesssim 1000$ and mass loss, the ratio is $\lesssim$ 6\%.  }

Our deep HST narrow-band imaging places the strongest constraint yet on the SFR$_{\rm{PopIII}}$ in high-redshift galaxies. Although we have not detected the He II emission line in IOK-1 at more than 1.2$\sigma$, this limit suggests that detection of He II emission from Pop III stars may be already within the reach of current facilities. Future facilities, especially JWST and GSMT, will be likely to probe down to the $\gtrsim 3\sigma$ level and directly detect the signatures of the earliest star formation in the universe.

 We thank the anonymous referee for insightful comments which
have significantly improved the Letter. 
We thank Stefano Casertano, Daniel Schaerer, and Haojing Yan for useful discussions.  
Support for this work was provided by NASA through grant HST-GO-11587  from the Space Telescope Science Institute, which is operated by AURA, Inc., under NASA contract NAS5-26555. ZC, XF, LJ, FB and IM acknowledge support from NSF grant AST 08-06861, and a David and Lucile Packard Fellowship. AIZ acknowledges support from the NSF through grant AST-0908280 and from NASA through grant NNX10AD47G. 
   
%
%

%
%
\paragraph{}



\clearpage
\setcounter{figure}{0}

\figurenum{1}
\begin{figure}[tbp]
\epsscale{1}
\label{fig:images}
\plottwo{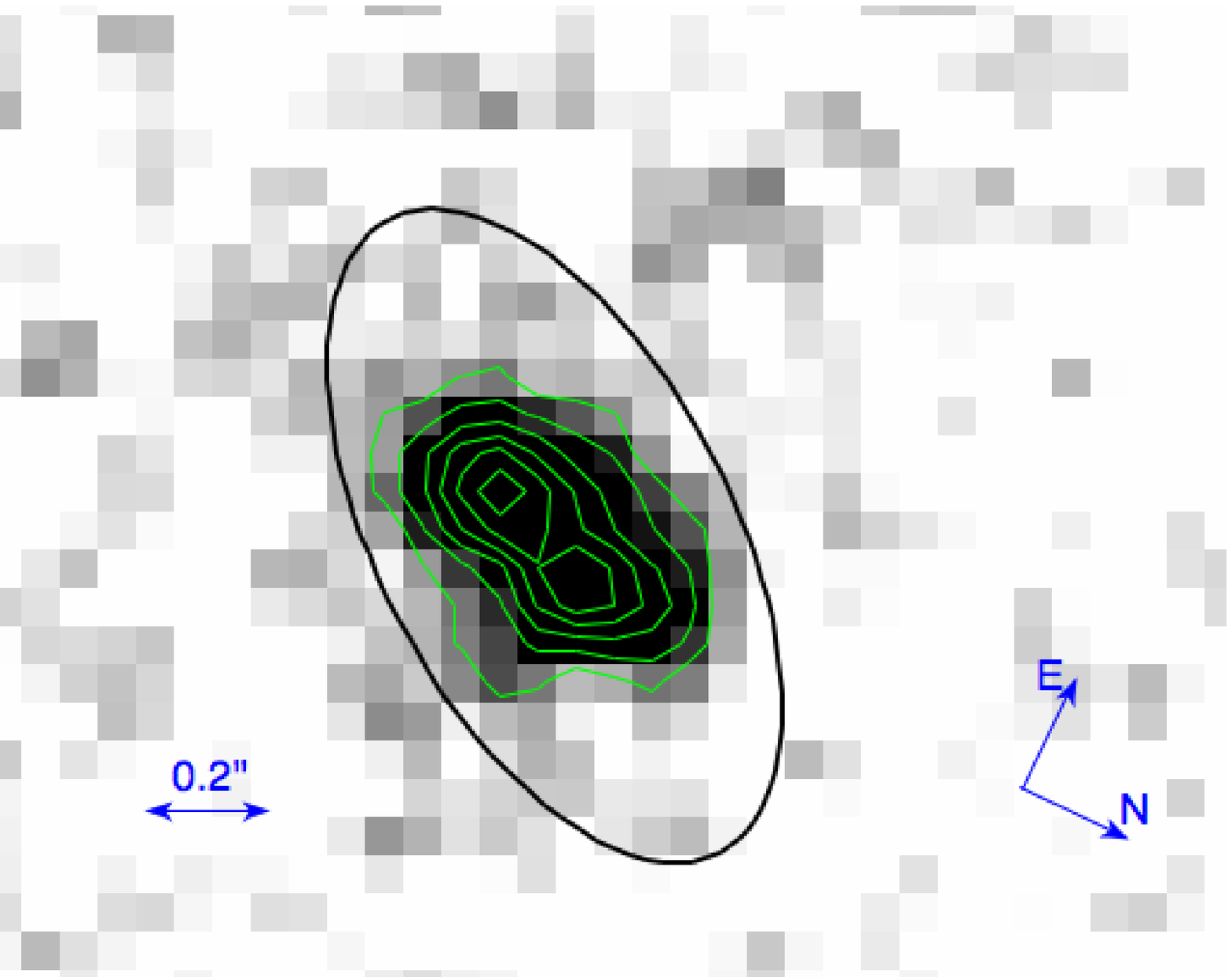}{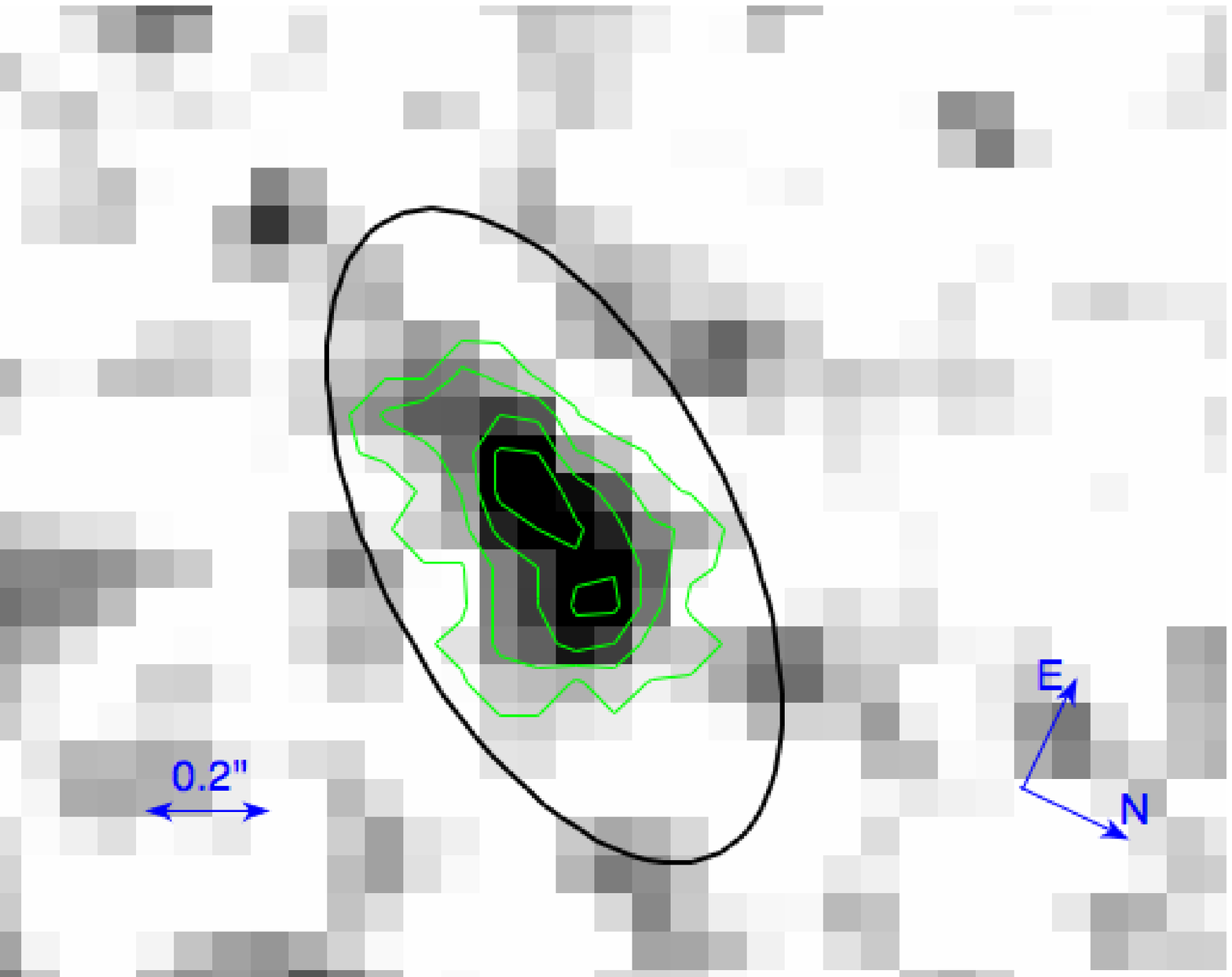}
\caption{High resolution images of galaxy IOK-1 in the F125W (left) and the F130N (right) bands, with contours spaced by 1.4 sky rms for the F125W image and 1 sky rms for the F130N image.  The black elliptical aperture is determined by the F130N image. 
IOK-1 is clearly resolved into two components separated by $\sim 0\farcs2$. Photometric analysis of the F130N image reveals a flux density excess of 1.2$\sigma$. }
\end{figure}

\figurenum{2}
\begin{figure}[tbp]
\epsscale{1}
\label{fig:testschemes}
\plotone{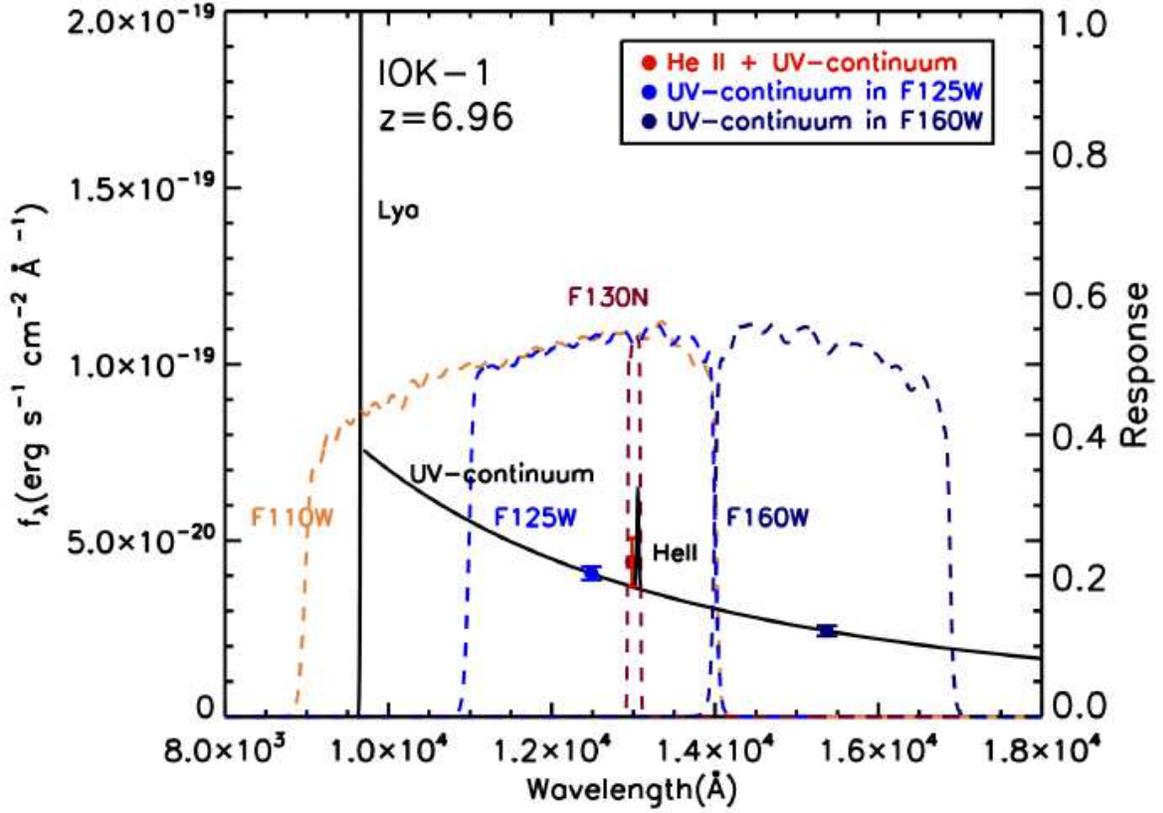}
\caption{Best-fit spectrum (black line) of galaxy IOK-1 with the total UV continuum, as well as the Ly$\alpha$ and He II emission lines. The filter response curves of F125W (blue dashed line), F160W (dark blue dashed line) and F130N (brown dashed line) are plotted. In addition, photometry in three different bands is overplotted at the effective wavelength of each filter.}
\end{figure}
\end{document}